\documentclass[article,aps,showpacs,amsfonts,epsf]{revtex4}

\usepackage{graphicx}

\renewcommand{\a}{\alpha}
\renewcommand{\k}{\kappa}

\newcommand{\be}{\begin{equation}}
\newcommand{\ee}{\end{equation}}
\newcommand{\bea}{\begin{eqnarray}}
\newcommand{\eea}{\end{eqnarray}}
\newcommand{\ba}{\begin{array}}
\newcommand{\ea}{\end{array}}
\def\J#1#2#3#4{{#1} {\bf #2}, #3 (#4)}
\def\PRD{Phys. Rev. D}
\def\PR{Phys. Rev.}
\def\PRL{Phys. Rev. Lett.}
\def\PTP{Prog. Theor. Phys.}

\def\APL{Ann. Phys. (Leipzig)}

\def\JMP{J. Math. Phys.}

\def\CQG{Class. Quantum Grav.}

\def\GRG{Gen. Relativ. Grav.}
\def\PLA{Phys. Lett. A}

\def\PL{Phys. Lett.}

\begin{document}
\draft
\title{Are known maximal extensions of the Kerr and
Kerr-Newman spacetimes physically meaningful and analytic?}

\author{H. Garc\'ia-Compe\'an and V.~S.~Manko}
\address{Departamento de F\'\i sica, Centro de Investigaci\'on y de
Estudios Avanzados del IPN, A.P. 14-740, 07000 M\'exico D.F.,
Mexico}

\begin{abstract}
In this paper we argue that the well-known maximal extensions of
the Kerr and Kerr-Newman spacetimes characterized by a specific
gluing (on disks) of two asymptotically flat regions with ADM
masses of opposite signs are physically inconsistent and actually
non-analytic. We also discover a correct geometrical
interpretation of the surface $r=0$, $t={\rm const}$ -- a {\it
dicone} in the case of the Kerr solution and a more sophisticated
surface of non-zero Gaussian curvature in the case of the
Kerr-Newman solution -- which suggests that the problem of
constructing the maximal analytic extensions for these stationary
spacetimes is likely to be performed within the models with only
one asymptotically flat region, in which case a smooth crossing of
the ring singularity becomes possible, for instance, after
carrying out an appropriate transformation of the radial
coordinate. \end{abstract}


\maketitle


\section{Introduction}
\label{intro}

Unlike the maximal analytic extensions (MAEs) of the Schwarzschild
\cite{Sch,Kru} and Reissner-Nordstr\"om \cite{Rei,Nor,GBr,Car1}
metrics in which the range of the radial coordinate $r$ is
restricted to non-negative values only, the MAEs of the Kerr
\cite{Ker,BLi} and Kerr-Newman (KN) \cite{NCC,Car2} spacetimes
involve both positive and negative values of $r$. This seems
needed because the intrinsic curvature singularity located at
$r=0$, $\theta=\pi/2$ in the equatorial plane of each of the
latter two spacetimes has topology of a ring and the extension
within the same spacetime, through the disk enclosed by the ring
singularity, is known to be not $C^1$ \cite{GPo}. The procedure of
continuation of $r$ into negative values, which involves gluing of
the regions with positive and negative $r$ on the disks $r=0$, is
well described in the classical books on general relativity
\cite{HEl,Wal} and leads to the appearance of the second
asymptotically flat region of negative ADM \cite{ADM} mass,
provided the mass of the first asymptotically flat region is
positive definite. The very fact that the same singularity may
look as having positive or negative mass depending on a particular
asymptotically flat region in which an observer is situated seems
rather unphysical, but quite surprisingly this has never been
questioned or objected in the literature. An additional not quite
desirable feature of the known MAEs of the Kerr and KN solutions
is that in the static limit they do not reduce straightforwardly
to the MAEs of static spacetimes with only one asymptotically flat
region. Moreover, although it is generally thought (but has never
been proved!) that the specific gluing of two asymptotically flat
regions employed in those MAEs is smooth, a recent study of the
Kerr and KN solutions endowed with negative mass \cite{Man1,Man2}
has revealed, however, that this might be not true. Indeed, it
turns out that the curvature singularities in the negative-mass
case are {\it massless}, being located {\it outside} the mass
distributions, which makes them clearly distinctive from the
singularities in the positive-mass case, thus raising doubts not
only about the analyticity of the corresponding MAEs on the disks
joining the regions of positive and negative ADM masses but also
about the disk interpretation of the surface $r=0$ itself. In the
present paper we will discuss the intrinsic physical and
mathematical inconsistencies of the MAEs from
Refs.~\cite{BLi,Car2}, and a surprising outcome of that discussion
will be discovery of a correct geometry of the surface $r=0$ -- a
{\it dicone} in the case of the Kerr solution, and a specific {\it
two-surface of positive and negative Gaussian curvature} in the
case of the KN solution -- thus showing that the old well-known
interpretation of the surface $r=0$ as a disk is wrong, as well as
the whole scheme of gluing on disks of two asymptotically flat
regions employed in the known MAEs. Our findings strongly favor
the construction of the maximally extended Kerr and KN solutions
within the framework of the models with only one asymptotically
flat region, and we demonstrate in particular that, in an
appropriate coordinate system, the Kerr and KN spacetimes allow
one to pass analytically from one hemisphere to another through
the part of the equatorial plane encircled by the ring
singularity, which means that the ring singularity is in fact
smoothly traversable.

\section{Physical and mathematical inconsistences of the known MAEs}
\label{sec:1}

The KN metric in the Boyer-Lindquist coordinates
$(r,\theta,\varphi,t)$ reads as
\be d s^2=\Sigma\left(\frac{d r^2}{\Delta}+d\theta^2\right)
+\frac{\sin^2\theta}{\Sigma}[a d
t-(r^2+a^2)d\varphi]^2-\frac{\Delta}{\Sigma}(d t-a\sin^2\theta
d\varphi)^2, \label{KN_m} \ee
with
\be \Delta=r^2-2Mr+a^2+Q^2, \quad \Sigma=r^2+a^2\cos^2\theta,
\label{dd} \ee
where $M$, $a$ and $Q$ are the mass, rotational and charge
parameters, respectively, and the coordinate range is: $0\le
r<\infty$, $0\le\theta\le\pi$, $0\le\varphi\le2\pi$,
$-\infty<t<\infty$. When $Q=0$, Eqs.~(\ref{KN_m}) and (\ref{dd})
define the Kerr metric. The curvature singularity is determined by
$\Sigma=0$, i.e., it is located at $r=0$, $\theta=\pi/2$ in the
equatorial plane, the singularity for $M\ne0$, $a\ne0$ being
ring-like rather than point-like, which allows one to extend the
KN metric into the negative values of the radial coordinate $r$.
Then a ``standard'' extension of the region~I ($0\le r<\infty$),
with $M>0$, based on the interpretation of the surface $r=0$ as a
disk (or a cut) \cite{Isr}, consists in letting $r$ take negative
values within a different KN sheet ($-\infty<r\le 0$, $M>0$), thus
giving rise to the region~II which is equivalent, in view of the
invariance of (\ref{KN_m}) under the discrete transformation
$r\to-r$, $M\to-M$, with the KN spacetime corresponding to $M<0$,
$r\ge 0$ and characterized by negative ADM mass. In the MAEs from
\cite{BLi,Car2}, employing a specific gluing of the regions I and
II on the ``disks'' (see Fig.~1), it is assumed implicitly that
the ring singularity present in the Kerr and KN solutions with
positive mass is the same as in the negative-mass case, so that in
order to formally pass from the region~I to region~II and vice
versa one uses the sole singularity, the latter being seen as
having positive mass in the region~I, and having negative mass in
the region~II. This situation, when there exists the ``other
side'' where the singularity's mass changes its sign, really looks
surpassing any logic and imagination, and it is not really
surprising that up to date the question of how the singularity's
mass may have simultaneously two opposite values has not been
clarified in the literature.

\begin{figure}
  \centering
    \includegraphics[width=90mm]{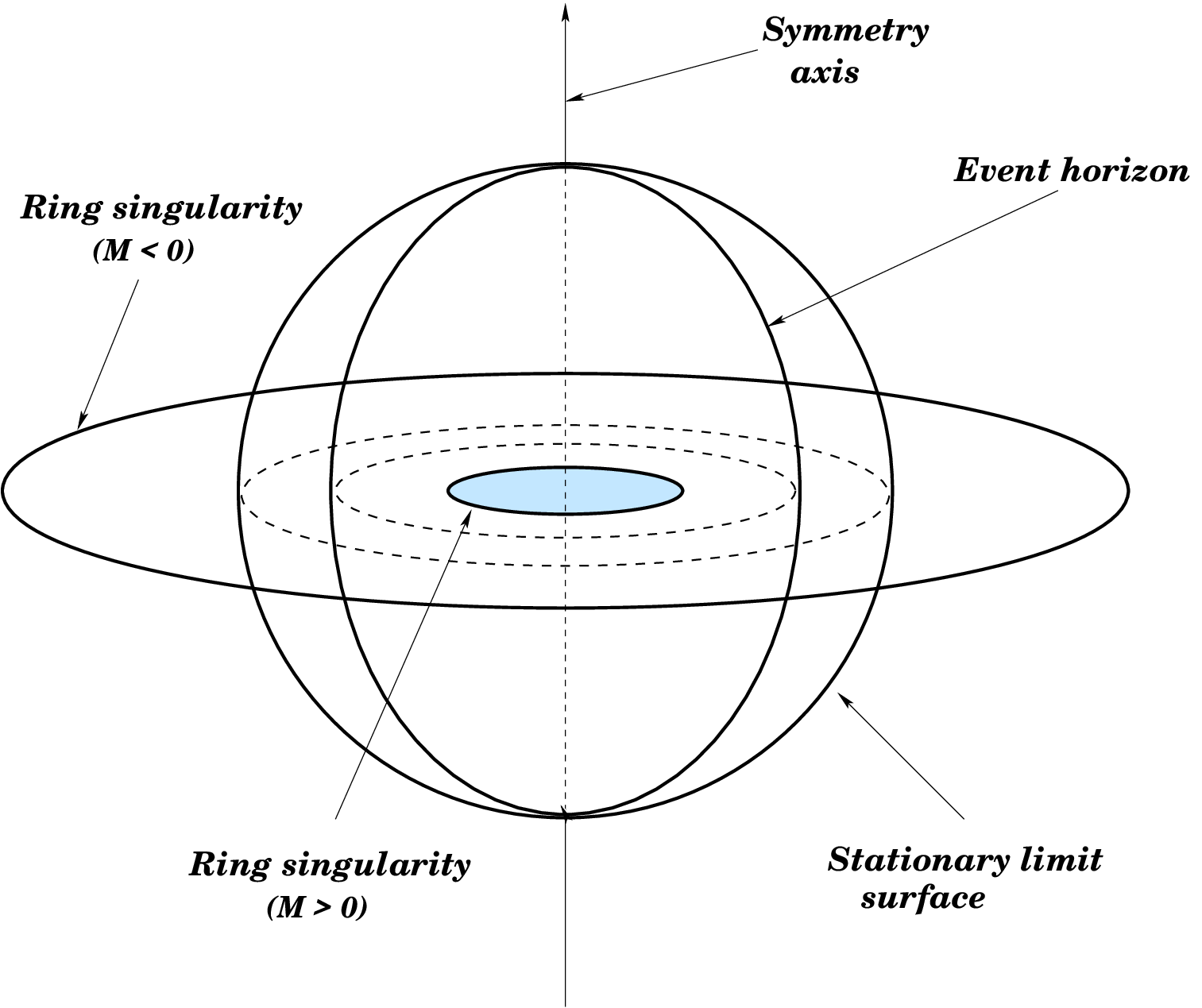}
  \caption{Standard scheme of gluing the $r>0$ and $r<0$ sheets of the Kerr
  (or KN) solution on the surfaces $r=0$ when the latter surfaces are
  interpreted as equatorial disks: the upper side of each disk
  must be identified with the bottom side of the other disk.
  Note that this special ``upper-bottom'' identification is required
  to ensure the continuity of the analytic extension in the polar
  angle $\theta$. Note also that the ring singularities are
  contained {\it inside} the surfaces $r={\rm const}$ in both the
  regions I and II, i.e. independently of the sign of $r$.}
  \label{fig1}
\end{figure}

Now, as it follows from the above (see also \cite{GPo}), the known
MAEs of the Kerr and KN metrics can be envisaged, roughly
speaking, as a unification of two physically different manifolds:
(i) the usual black-hole or hyperextreme spacetime ($M>0$,
$r\ge0$) and (ii) a spacetime created by negative mass ($M<0$,
$r\ge 0$, which is equivalent to region~II: $M>0$, $r\le 0$). We
would like to emphasize that the appearance of the negative-$r$
asymptotically flat region II, as well as the need for making use
of the gluing procedure in order to combine the regions I and II
in one manifold, is a direct consequence of the disk
interpretation of the surface $r=0$ because the region I in that
case is unable to have any subregion of $r<0$. At the same time, a
crucial point to observe here is that the ring singularity $r=0$,
$\theta=\pi/2$ is not the same object in the $M>0$ and $M<0$
cases. This follows directly from the recent study
\cite{Man1,Man2} of the Kerr and KN solutions with negative mass:
if for instance $M^2>a^2+Q^2$, the singularity in the
positive-mass case is massive and lies inside the horizon
\cite{NJa}, while in the $M<0$, $r\ge0$ case it is massless and
lies off the symmetry axis outside the stationary limit surface
\footnote{Location of the singularity of the negative mass KN
solution at $\rho>0$, $z=0$ was also mentioned by Meinel
\cite{Mei} (see a footnote on p.~10 of his paper).}; the only
common feature shared by the two singularities is their $S^1$
topology. In other words, for $M\ne0$, the limit $r\to0$ leads to
two different singular rings corresponding to $M>0$ and $M<0$ (see
Fig.~2), thus spoiling the analyticity property of the standard
extensions of the Kerr and KN solutions which fail to be smooth on
the disks joining the two asymptotically flat regions. Mention
also that in the absence of rotation ($a=0$), when
Eqs.~(\ref{KN_m}) and (\ref{dd}) define the Reissner-Nordstr\"om
metric, the curvature singularity in the $M>0$ case is point-like
while in the $M<0$ case it has topology of a sphere, which means
that the ``standard'' MAE of the KN solution does not admit a
mathematically correct electrostatic limit.

\begin{figure}
  \centering
    \includegraphics[width=90mm]{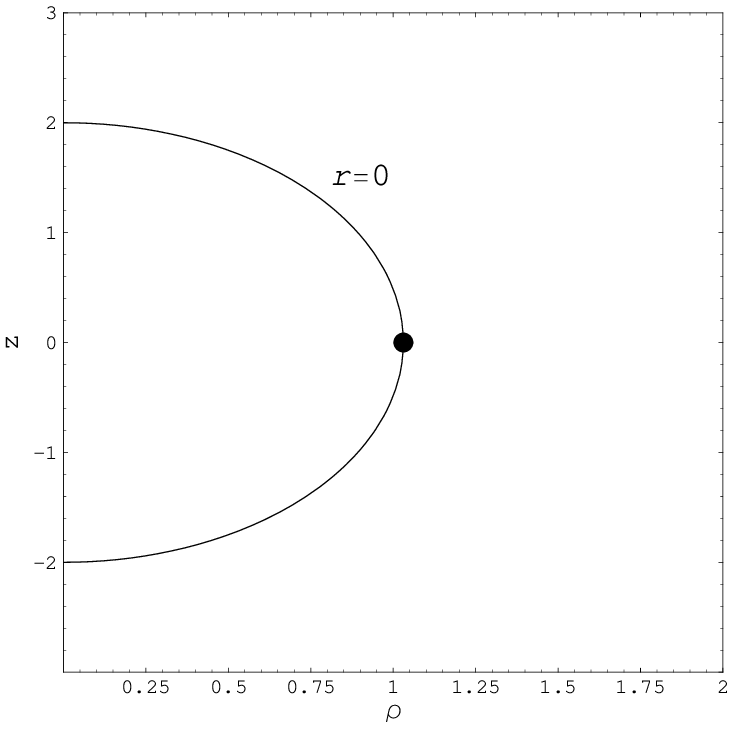}
  \caption{Ring
singularities of the KN solution in the $M>0$ and $M<0$ cases. The
correct location of the ring singularities with respect to each
other reveals itself after a rigorous derivation of the KN
solution in the Weyl-Papapetrou cylindrical coordinates via a
solution generating technique.}
  \label{fig1}
\end{figure}

Apparently, the main physical inconsistency of the known MAEs of
the Kerr and KN solutions is the non-correspondence of the
asymptotics of region~II to the positive-mass singularity as a
source for that region artificially introduced instead of a
genuine source of negative mass. In this respect one should bear
in mind that the region~II will be always characterized
asymptotically by the same ADM mass independently of which source
one puts at its $r=0$; however, the source's mass must be the same
as measured asymptotically if one wants to obtain a physically
consistent model. On the other hand, the main mathematical problem
of the known MAEs is the non-analyticity of the gluing procedure
on the ``disks'' $r=0$ encircled by the ring singularities (some
problems with the differentiability of the Kerr and KN spacetimes
at $r=0$ were already mentioned in \cite{Pun}). As a matter of
fact, the general belief in the analyticity of the latter
procedure on the disks joining the regions I and II is just a
wishful thinking which has never received a rigorous proof in the
literature. Thus, for instance, Boyer and Lindquist themselves
only remark that ``a continuation to negative $r$ values is
permissible because the Kerr metric remains regular at $r = 0$
(provided $\theta\ne\pi/2$)'' (p.~270 of \cite{BLi}), not touching
the issue of smoothness of the gluing procedure itself; however,
as will be seen in the next Section, even the aforementioned claim
on the regularity of $r=0$ turns out to be misleading because of
the presence on that surface of two singular cusps.

\section{Towards the construction of correct MAEs}

At first glance it seems that the physical inconsistency inherent
in the aforementioned MAEs can be remedied by only a slight
modification of the extension scheme: while the region~I
corresponds, as in Refs.~\cite{BLi,Car2}, to positive values of
$r$ and $M$, the region~II should involve negative values of both
$r$ and $M$. Consequently, the newly defined region~II, with $r\le
0$, $M<0$, will be identical with the region~I (again by virtue of
invariance of the metric~(\ref{KN_m}), (\ref{dd}) under the
discrete transformation $r\to-r$, $M\to-M$) and of course will
have the same asymptotics as region~I. The MAE performed in this
way looks physically consistent because the resulting extended
manifold does not create any conflict of singularities, which are
this time the same curvature ring singularity endowed with
positive mass, and both regions~I and II are described by the same
ADM mass coinciding with the singularity's mass. Certain support
in favor of this way of continuing $r$ into negative values in the
Kerr and KN solutions comes from Carter's MAE of KN spacetime
itself: in the case $M=0$, this MAE seems to become physically
reasonable and coincides with the above construction, being
composed of two identical asymptotically flat regions. The KN
solution in this case represents the exterior field of a massless
charged magnetic dipole, and the metric is still stationary due to
a specific frame-dragging effect \cite{Das,Bon}, with
$g_{t\varphi}= aQ^2\sin^2\theta/(r^2+a^2\cos^2\theta)$ and a
zero-mass ring singularity. However, the use of the same procedure
of attaching the region~II to region~I as described in detail by
Hawking and Ellis for the Kerr MAE \cite{HEl}, which consists in
identifying the top side of the first disk with the bottom side of
the second one (and vice versa), in the case of two identical KN
spacetimes is in fact similar to the problem of analytically
extending a single KN metric (with $M>0$, $r\ge0$) through the
disk encircled by the ring singularity, and the latter extension
is commonly thought to be not $C^1$ (see, e.g., \cite{GPo}). Thus,
one arrives at the following delicate situation: the known MAEs of
the Kerr and KN solutions are not satisfactory both physically and
mathematically, but a rectified extension procedure within the
framework of the same philosophy of two asymptotically flat
regions, being physically consistent, leads to the mathematical
problem that long ago had already forced the researchers to
abandon the philosophy of the MAE with only one asymptotically
flat region. We think that the only plausible way to resolve this
seeming contradiction would be demonstrating that the existing
evidence against the traversability of the ring singularity in a
single KN spacetime is not reliable.

Actually, the question of whether or not the KN spacetime can be
smoothly continued through the ring singularity from one
hemisphere to another is of paramount importance, being a key
point for the construction of a correct MAE, and until now this
question had received a negative answer in the literature
\cite{GPo,Isr}. The usual ``non-traversability'' argument is the
following: the disk inside the ring singularity is assumed to be
described by the surface $r=0$, and across that surface there are
discontinuities in the first derivatives of the metric tensor
components that can be associated with the presence of the matter
sources. Then one ``naturally'' concludes that the analytic
continuation through the disk is impossible. Nonetheless, as was
already observed in \cite{Pun}, there exist certain problems of
differentiability when one analyzes the Kerr geometry in the
vicinity of $r=0$. In our opinion, these problems could hardly
arise if the surface $r=0$ were indeed a disk, which in turn
raises a legitimate question of establishing a correct geometrical
interpretation of the latter surface. In this relation, below we
are going to demonstrate (i) that for $M\ne0$, the surface $r=0$
does not actually describe a disk; as a result, it does not even
lie in the equatorial plane (the ring singularity being just an
intersection of $r=0$ with the equatorial plane), and (ii) that
after an appropriate coordinate transformation, a smooth extension
across the ring singularity becomes possible.

\begin{figure}
  \centering
    \includegraphics[width=90mm]{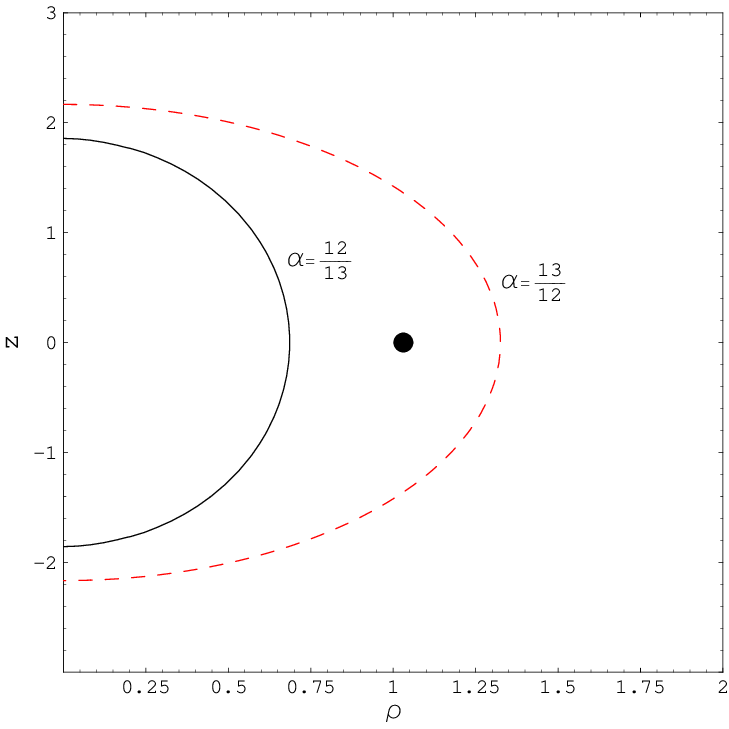}
  \caption{The form
of the surface $r=0$ in cylindrical coordinates demonstrating that
the Boyer-Lindquist coordinates with $r\ge0$ do not cover the
whole space.}
  \label{fig2}
\end{figure}

To make our demonstration more visual, we will first consider the
KN solution with negative mass, in which case the singular ring is
also ``non-traversable'' in the Boyer-Lindquist coordinates with
$r\ge0$. The consideration of this case will permit us in
particular to make use of a known coordinate transformation for
establishing the precise geometry of the surface $r=0$. According
to \cite{Man2}, the KN spacetime with negative mass can be
conveniently tackled in the usual Weyl-Papapetrou cylindrical
coordinates $\rho$ and $z$ ($\rho\ge 0$, $-\infty<z<\infty$) which
are related to the Boyer-Lindquist coordinates ($r,\theta$) via
the formulas
\bea r&=&M+\frac{1}{2}(r_++r_-), \quad
\cos\theta=\frac{1}{2\k}(r_+-r_-), \nonumber\\
r_\pm&=&\sqrt{\rho^2+(z\pm\k)^2}, \quad \k=\sqrt{M^2-a^2-Q^2}.
\label{cyl} \eea
With the aid of (\ref{cyl}), in Fig.~3 we have plotted the
hypersurface $r=0$ in cylindrical coordinates for the particular
parameter choice $M=-2$, $a=1$, $Q=0.25$, and one can see that it
can be {\it tentatively} interpreted as a spheroid (a more precise
interpretation will be given later on) with the poles $z=\pm 2$,
$\rho=0$ and the ring singularity $z=0$, $\rho\approx1.031$ as its
equator.

For the general negative-mass case we find from (\ref{cyl}) that
the surface $r=0$ in the $(\rho,z)$ coordinate space is defined by
the equation
\be \frac{\rho^2}{a^2+Q^2}+\frac{z^2}{M^2}=1, \label{surf} \ee
i.e., it represents a spheroid with the poles $z=\mp M$, $\rho=0$
and the equatorial radius $\sqrt{a^2+Q^2}$, the ring singularity
$z=0$, $\rho=\sqrt{a^2+Q^2}$ lying on its equator. In the
Boyer-Lindquist coordinates with $r\ge0$ the spheroid is therefore
removed from the general manifold, which, on the one hand,
explains the existence of the aforementioned discontinuities in
the first derivatives of the metric tensor on the two sides of the
surface $r=0$, and, on the other hand, disproves the known
interpretation \cite{Isr} of the latter surface as ``an equatorial
disk'' or a cut in the equatorial plane. The spheroid degenerates
to an infinitesimally thin disk only in the case $M=0$ when the
discontinuities disappear, so that, when $M\ne 0$, $r\ge0$, the
Boyer-Lindquist coordinates cannot for example describe the
genuine part of the equatorial plane encircled by the ring
singularity as it does not coincide with the surface $r=0$ and
lies inside the spheroid. Remarkably, the latter problem does not
occur after rewriting the KN metric in cylindrical coordinates
when one becomes capable to reach analytically the part
$0<\rho<\sqrt{a^2+Q^2}$ of the equatorial plane and then smoothly
cross the latter in the absence of discontinuities in the
derivatives of the metric tensor. In this case the ring
singularity turns out to be perfectly well traversable because no
artificial cut in the form of a disk or a spheroid arises in
cylindrical coordinates, the latter fully covering the region
$\rho\ge 0$, $-\infty<z<\infty$, contrary to the Boyer-Lindquist
coordinates with $r\ge0$. Moreover, in the equatorial plane the
metric coefficient $g_{tt}$ of the KN metric with $M<0$,
$M^2>a^2+Q^2$ takes the form
\be g_{tt}=\frac{\rho^2-a^2}{M+\sqrt{\rho^2+M^2-a^2-Q^2}},
\label{gtt} \ee
whence it follows that at the part of the (genuine) equatorial
plane encircled by the ring singularity the KN metric is sensitive
to the sign of mass, thus excluding the possibility of gluing
there spacetimes characterized by masses of opposite signs.

By introducing instead of $r$ a new radial coordinate $\tilde r$
via the formula
\be \tilde r=\a M+\frac{1}{2}(r_++r_-), \quad \a={\rm const},
\label{rt} \ee
with $\tilde r\ge0$, it is easy to see that an appropriate choice
of the parameter $\a$ permits one to get a better description of
the KN solution than that given by the usual Boyer-Lindquist
coordinates. In Fig.~4 we have plotted the hypersurface $\tilde
r=0$ for two particular values of $\a$ (the values of $M$, $a$ and
$Q$ are the same as in Fig.~3); the case $\a=12/13$ gives an
extension of the Boyer-Lindquist coordinates thus making the ring
singularity partially traversable, while the choice $\a=13/12$
leads to a larger restriction of the KN manifold than in Fig.~3
(remember that $M<0$). In general, for all $\a M>-\k$, the new
coordinates ($\tilde r,\theta$) may be considered as extensions of
the cylindrical coordinates ($\rho,z$), and then the ring
singularity becomes fully traversable.

\begin{figure}
  \centering
    \includegraphics[width=90mm]{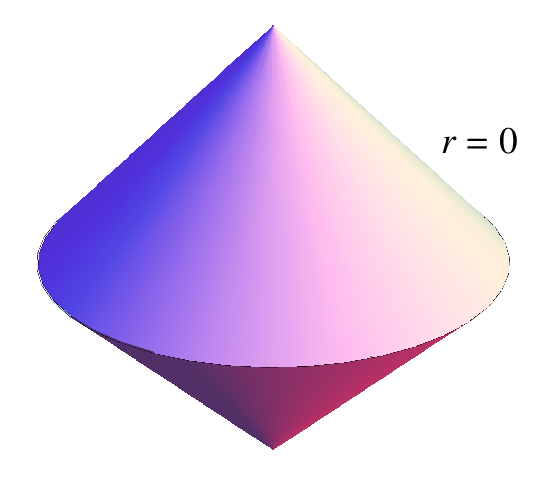}
  \caption{The
surface $\tilde r=0$ for two particular values of the parameter
$\a$. The coordinate transformation with $\a=12/13$ (solid line)
makes the ring singularity partially traversable.}
  \label{fig3}
\end{figure}

To have a better idea about the geometry of the surface $r=0$,
$t={\rm const}$, defined by (\ref{surf}) in the negative-mass case
of the KN solution, we have calculated with the aid of the
computer analytical program RICCI \cite{Agu} the Gaussian
curvature $K=R/2$ of the corresponding two-dimensional metric
\be ds_{(2)}^2=\frac{a^2z^2}{M^2(M^2-z^2)}dz^2
+\frac{(M^2-z^2)[(a^2+Q^2)z^2-M^2Q^2]}{M^2z^2}d\varphi^2,
\label{rz2} \ee
yielding
\be K=\frac{M^4Q^2[(a^2+Q^2)z^2(4M^2-z^2)-3M^4Q^2]}
{a^2z^4[(a^2+Q^2)z^2-M^2Q^2]^2}. \label{Kz} \ee

It is easy to see from the above formula that the sign of $K$
depends on $z$ in the following way: for $z_0^2<z^2<M^2$, where
\be z_0=\pm|M|\sqrt{2-\sqrt{\frac{4a^2+Q^2}{a^2+Q^2}}}, \label{z0}
\ee
$K$ is positive, while for $0<z^2<z_0^2$, $K$ is negative.
Moreover, $K\to +\infty$ in the limit $z^2\to M^2Q^2/(a^2+Q^2)$,
and $K\to -\infty$ when $z\to 0$. Therefore, it is evident that
for a non-vanishing $Q$ the surface $r=0$ of the KN solution with
negative mass is not a disk, but a specific two-surface of
positive and negative Gaussian curvature. Although, as follows
from our analysis, it is not actually a spheroid, it does of
course represent a surface of revolution.

The limit $Q=0$ in the formula (\ref{Kz}) is of special interest
and hence needs to be considered in more detail. In this limit the
KN solution reduces to the Kerr space and the Gaussian curvature
$K$ of the two-surface $r=0$, $t={\rm const}$ becomes equal to
zero, which at first glance seems to be in agreement with the
already known analogous result for the Kerr solution with positive
mass (see, e.g., Ref.~\cite{ONe}). However, whereas in \cite{ONe}
the zero-value Gaussian curvature is used as an argument in favor
of a disk-type form of the surface $r=0$ (in the absence of any
preliminary picture of the latter), it is difficult to accept such
an interpretation having at hand a precise equation (\ref{surf})
of that surface, which does not cease to represent a spheroid when
$Q=0$, together with a particular image from Fig.~3, so that we
definitely must look for a different explanation of vanishing $K$
in the Kerr case. Fortunately, there exists a simple alternative
for an equatorially symmetric closed two-surface to have zero
Gaussian curvature and at the same time not to be a disk -- it is
a {\it dicone}, i.e. a surface formed by two identical cones
joined together by their bases (see Fig.~5). This new
interpretation of Kerr's $r=0$ surface is supported by all our
previous considerations and is congruent with our interpretation
of the respective $r=0$ surface of the KN solution, so that in
principle one may view the latter surface as Kerr's dicone
perturbed by the electromagnetic field.

\begin{figure}
  \centering
    \includegraphics[width=75mm]{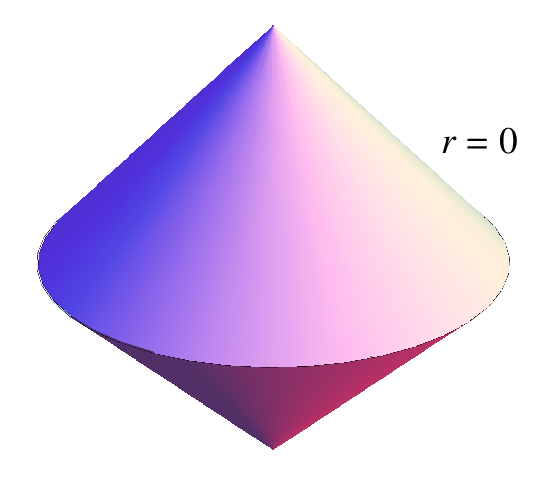}
  \caption{The surface $r=0$, $t={\rm const}$ of the Kerr solution -- a dicone.
  The ring singularity $r=0$, $\theta=\pi/2$ resides on the equator of the dicone.}
  \label{fig3}
\end{figure}

In order to justify in a rigorous way the dicone geometry of the
surface $r=0$ in the Kerr case, it is likely first to supplement
the two-metric (\ref{rz2}) with its $(\rho,\varphi)$
representation. Then, from (\ref{surf}) and (\ref{rz2}) we get
\be ds_{(2)}^2=\left(1-\frac{Q^2}{a^2+Q^2}\right)d\rho^2
+\left(1-\frac{Q^2}{\rho^2-a^2-Q^2}\right)\rho^2d\varphi^2,
\label{rf2} \ee
and a simple check shows that the metric (\ref{rf2}) provides us
with the same information about the Gaussian curvature $K$ of the
surface $r=0$ as the metric (\ref{rz2}) before: $K\ne0$ in the KN
case and $K=0$ in the Kerr case. Now, by setting $Q=0$ in
(\ref{rf2}), we arrive in the latter case at
\be ds_{(2)}^2=d\rho^2 +\rho^2d\varphi^2, \label{flat2} \ee
which is a familiar two-metric determining three different flat
surfaces, and one only needs to make a correct choice among them
for establishing the non-local geometry of Kerr's $r=0$ surface.
Remarkably, such a choice does not represent any difficulty thanks
to the surface's equation (\ref{surf}) and the metric (\ref{rz2})
which jointly discard a disk and a cylinder as possible geometries
for the surface $r=0$. Indeed, it is obvious that the disk
geometry must be excluded because the coordinate $z$ in
(\ref{rz2}) runs all values from the interval $[-|M|,+|M|]$, while
$z$ in the case of a disk must take only one particular value; on
the other hand, the surface $r=0$ cannot be a cylinder due to the
presence in the metric (\ref{rz2}) of two singular points $z=\mp
M$ locating, as it follows from (\ref{surf}), on the symmetry axis
$(\rho=0)$. The singularity structure of the metric (\ref{rz2})
also reveals that the implementation of the remaining (conic)
option in the Kerr case involves two identical cones combined in a
dicone, taking into account that the Kerr solution is symmetric
with respect to the equatorial ($z=0$) plane and that $z\le|M|$.
The opening angle $\a_0$ of Kerr's dicone is defined by the
formula
\be \tan\frac{\a_0}{2}=\frac{|a|}{|M|}, \label{oa} \ee
so that in the limit of vanishing mass, $\a_0$ tends to $\pi$ and
the dicone degenerates to a disk.

Now we can discuss the case $M>0$ which turns out to be fully
analogous to the previous case of negative mass. Indeed, since the
non-negative values of the radial Boyer-Lindquist coordinate $r$
do not permit one to go beyond the ring singularity, an extension
of $r$ is needed to describe geometry inside the singular ring.
Like in the $M<0$ case, the hypersurface $r=0$ for $M>0$ does not
represent a disk, being a specific surface of rotation, and the
space enclosed by the latter is withdrawn from the general KN
manifold. It is clear that the extension of $r$ into negative
values with the aid of the scheme worked out in the papers
\cite{BLi,Car2} does not affect the interior of the closed surface
$r=0$, $t={\rm const}$ because that scheme treats the latter
surface as a disk and hence needs a second copy of the KN
spacetime for the negative values of $r$, together with a
procedure of gluing the two manifolds on the ``disks".

Bearing in mind that the geometry of the KN solution with positive
mass must coincide in the vicinity of $r=0$ with the negative-mass
KN geometry in view of the aforementioned invariance of the KN
metric under the discrete transformation $r\to-r$, $M\to-M$, we
have calculated, directly in the Boyer-Lindquist coordinates, the
Gaussian curvature of the surface $r=0$, $t={\rm const}$ defined
by the line element
\be ds_{(2)}^2=a^2\cos^2\theta d\theta^2
+(a^2-Q^2\tan^2\theta)\sin^2\theta d\varphi^2, \label{BL2} \ee
the result being
\be K=\frac{Q^2[3a^2-(a^2+Q^2)\sin^2\theta(2+\sin^2\theta)]}
{a^2\cos^4\theta(a^2\cos^2\theta-Q^2\sin^2\theta)^2}. \label{KBL}
\ee

We leave it to the reader as a simple exercise to verify that the
formula (\ref{KBL}) provides one with exactly the same information
about the geometry of the surface $r=0$ of the positive-mass KN
solution as formula (\ref{Kz}) calculated in the negative-mass
case, and also that the analog of formula (\ref{z0}) determining
the change of curvature sign is the following expression for
$\theta_0$:
\be \sin^2\theta_0=\sqrt{\frac{4a^2+Q^2}{a^2+Q^2}}-1. \label{t0}
\ee

Formula (\ref{KBL}) for the Gaussian curvature $K$ apparently
refutes the existing interpretation of the surface $r=0$, $t={\rm
const}$ of the KN solution as a disk, and we only wonder why this
formula has not been considered earlier in the literature. Of
course, since both two-metrics (\ref{rz2}) and (\ref{BL2}) do not
distinguish among positive and negative $M$, their $Q=0$
specializations (Kerr case) describe the same zero-$K$ geometry
that has been already discussed -- a dicone. Mention in this
connection that the surface $r=0$ has two cusps not only in the
Kerr case, where the singularities are associated with the
dicone's apices, but in the KN case too, as it follows from the
singularity structure of the two-metric (\ref{rz2}).

It is clear that the interior of the surface $r=0$ can be easily
incorporated, at least partially, into the general manifold (thus
making the ring singularity traversable for the geodesics) for
instance by introducing a new radial coordinate $\tilde r$ which
would extend in a natural way the KN spacetime inside the singular
ring, like this was earlier done in the negative-mass case. A
possible transformation may have the form
\be \tilde r=\sqrt{M^2+a^2}+\frac{1}{2}(r_++r_-), \quad
\cos\theta=\frac{1}{2\k}(r_+-r_-), \label{rnew} \ee
with $\tilde r\ge 0$, and $r_\pm$ and $\k$ earlier defined in
(\ref{cyl}). The main advantage of the above ($\tilde r,\theta$)
over the Boyer-Lindquist coordinates ($r,\theta$) is that $\tilde
r=0$ lies inside the surface $r=0$ and therefore does not contain
the ring singularity for any non-vanishing $a$. In the coordinates
(\ref{rnew}) the KN metric takes the form
\bea d s^2=\tilde\Sigma\left(\frac{d\tilde r^2}{\tilde\Delta}
+d\theta^2\right)+\frac{\sin^2\theta}{\tilde\Sigma}\{a d
t-[(\tilde r+M-\mu)^2+a^2]d\varphi\}^2 \nonumber\\
-\frac{\tilde\Delta}{\tilde\Sigma}(d t-a\sin^2\theta d\varphi)^2,
\label{KN_new} \eea
where
\be \tilde\Delta=(\tilde r-\mu)^2-\k^2, \quad \tilde\Sigma=(\tilde
r+M-\mu)^2+a^2\cos^2\theta, \quad \mu=\sqrt{M^2+a^2}, \label{dsn}
\ee
and now it has a traversable singular ring. The representation
obtained may be used as a starting point for elaborating, via the
usual procedures, a MAE involving non-negative $\tilde r$.
However, the construction of such MAE in explicit form, which
would probably not depend on a concrete nondisk geometry of
$r=0$,\footnote{We thank Malcolm MacCallum for drawing our
attention to this point.} goes beyond the scope of the present
paper. In Fig.~6 it is shown how the extension (with no gluing!)
across the surface $r=0$ must look like generically, and one can
see that it differs considerably from the extension in Fig.~1,
especially taking into account that the interior of $r=0$ may in
principle be covered by a certain interval of negative values of
the usual Boyer-Lindquist coordinate $r$ (as an alternative to
small positive values of $\tilde r$).

\begin{figure}
  \centering
    \includegraphics[width=75mm]{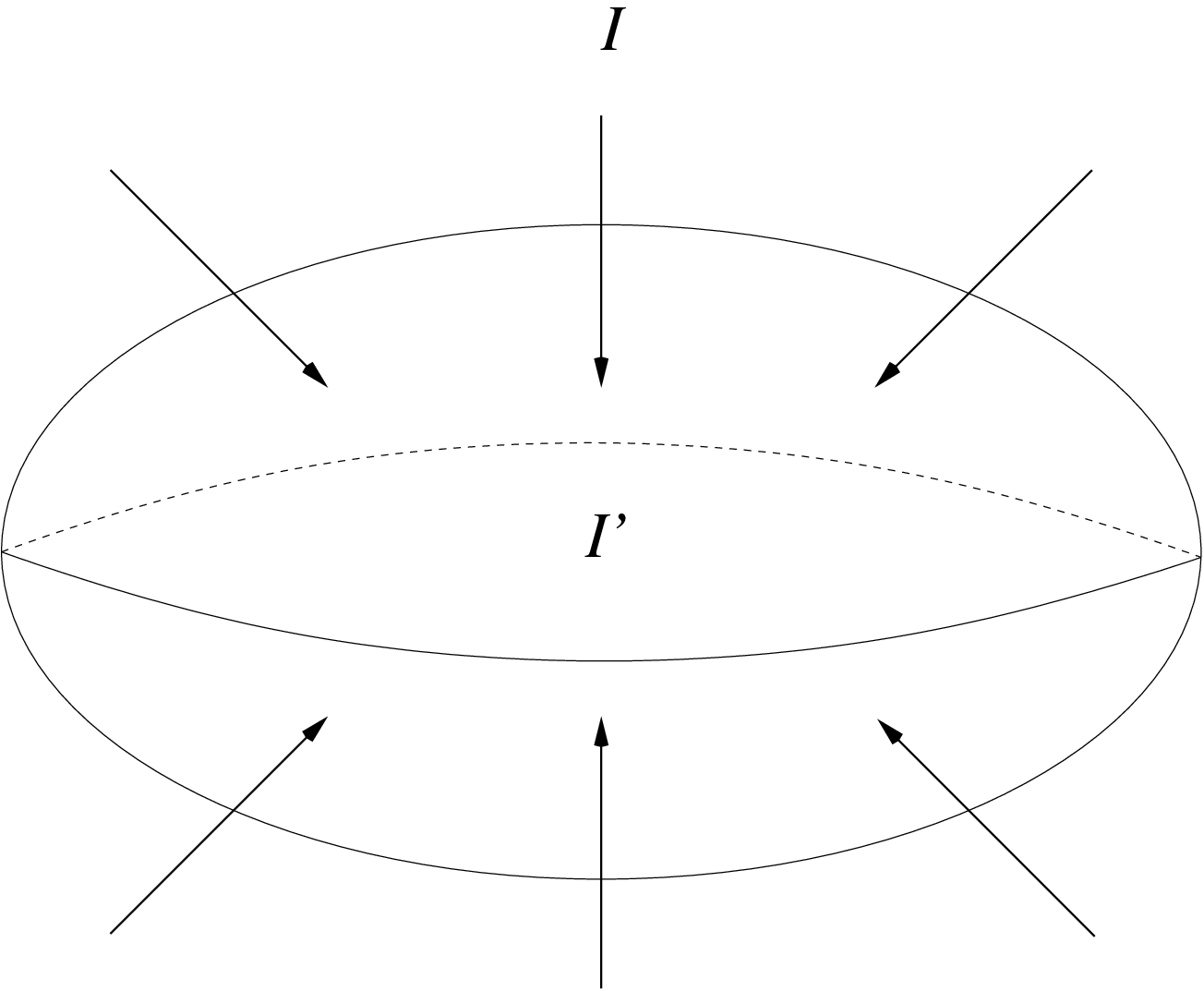}
  \caption{Extending correctly the Kerr (or KN) spacetime through the
  nondisk closed surface $r=0$: no gluing is needed,
  and the surfaces of constant negative $r$ (or positive $\tilde
  r$) from I' do not contain the ring
  singularity inside of them (unlike the analogous surfaces in the
  region II from Fig.~1).
  Note that the region I' (the surface $r=0$ being the boundary
  between the regions I and I') is
  not the same as the region II in Fig.~1; in particular, it does not
  contain the spatial infinity.}
  \label{fig3}
\end{figure}

It is worth noting that several analytically extended (by adding
the interior of the surface $r=0$ to the general manifold) KN {\it
identical} spacetimes could in principle be used for obtaining
MAEs of more complicated topologies by first making cuts in the
genuine equatorial disks inside the ring singularities (we recall
that these equatorial disks are not the $r=0$ surfaces) and then
gluing appropriately different sides of the cuts of different
copies. In this relation, a model involving two identical (and
previously extended) copies of the KN spacetime, which would
represent an improvement of the non-extended model with two
asymptotically flat regions mentioned by us earlier, gives a
possible example of such an exotic manifold in which the gluing
process does not cause problems and is mathematically justified.
Mention in conclusion that the extension of our analysis to the
case of the Kerr-Newman-(anti-)de Sitter solution \cite{GHa} is
straightforward.

\section*{Acknowledgments}

We wish to thank Eduardo Ruiz and Sergei Grudsky for interesting
conversations and helpful comments, and Malcolm MacCallum for
useful correspondence. V.M. is thankful to Norbert Van den Bergh,
Timothy Clifton, Kjell Rosquist and Lode Wylleman for a
comprehensive discussion of our results during the workshop
``Current Topics in Exact Solutions'' held at the Ghent University
in April 2014. This work was partially supported by Project~128761
from CONACyT of Mexico.




\end{document}